\documentclass[final,5p,times,twocolumn]{elsarticle}

\usepackage{amsmath}
\usepackage{amssymb}
\graphicspath{{plots/}}
\usepackage{xcolor}
\usepackage{algorithm}
\usepackage{algorithmicx}
\usepackage{algpseudocode}
\usepackage{hyperref}
\usepackage{url}
\usepackage{xcolor}
\usepackage{soul}

\journal{Parallel Computing}

\newcommand{\cmg}{\textrm{cm}^2\ \textrm{g}^{-1}}

\newcommand{\apj}{ApJ}
\newcommand{\apjl}{ApJL}
\newcommand{\apjs}{ApJS}
\newcommand{\nat}{Nature}

\newcommand{\mnras}{MNRAS}
\newcommand{\prl}{Phys. Rev. Lett.}
\newcommand{\prd}{Phys. Rev. D}

\begin{document}

\begin{frontmatter}

\title{Realistic Kilonova Up Close} 

\author[CCS3_LANL,MIT]{Alexandra Ruth Stewart\corref{alexa}}
\author[CCS3_LANL]{Li-Ta Lo}
\author[CCS7_LANL]{Oleg Korobkin}
\author[CCS2_LANL]{Irina Sagert}
\author[CCS7_LANL]{Julien Loiseau}
\author[CCS2_LANL]{Hyun Lim}
\author[CCS2_LANL,GWU]{Mark Alexander Kaltenborn}
\author[HPC_LANL]{Christopher Michael Mauney}
\author[CCS2_LANL]{Jonah Maxwell Miller}

\address[CCS2_LANL]{CCS-2, Computational Physics and Methods, Los Alamos National Laboratories, Los Alamos, NM 87545, USA}
\address[CCS3_LANL]{CCS-3, Information Sciences, Los Alamos National Laboratories, Los Alamos, NM 87545, USA}
\address[CCS7_LANL]{CCS-7, Applied Computer Science, Los Alamos National Laboratories, Los Alamos, NM 87545, USA}
\address[HPC_LANL]{HPC Environments, Los Alamos National Laboratories, Los Alamos, NM 87545, USA}
\address[MIT]{Department of Physics, Massachusetts Institute of Technology, Cambridge, MA 02139, USA}
\address[GWU]{Department of Physics, The George Washington University, Washington, DC 20052, USA}
\cortext[alexa]{stew@mit.edu}

\begin{abstract}
  Neutron star mergers are cosmic catastrophes that produce some of the most energetic observed phenomena: short gamma-ray bursts, gravitational wave signals, and kilonovae.
  The latter are optical transients, powered by radioactive nuclides which are synthesized when the neutron-rich ejecta of a disrupted neutron star undergoes decompression.
  We model this decompression phase using data from simulations of post-merger accretion disk winds.
  We use smoothed particle hydrodynamics with realistic nuclear heating to model the expansion over multiple scales, from initially
  several thousand km to billions of km.
  We then render a realistic image of a kilonova ejecta as it would appear for a nearby observer. 
  This is the first time such a visualization is performed using input from state-of-the-art accretion disk simulations, nuclear physics and atomic physics.   
  The volume rendering in our model computes an opacity transfer function on the basis of the physical opacity, varying significantly with the inhomogeneity of the neutron richness in the ejecta.
  Other physical quantities such as temperature or electron fraction can be visualized using an independent color transfer function.
  We discuss several difficulties with the ParaView application that we encountered during the visualization process, and give descriptions of our solutions and workarounds which could be used for future improvements. 

\end{abstract}


\end{frontmatter}

\section{Introduction}
\label{sec:Introduction}

In recent years, coalescences of binary neutron stars (BNS) and black hole (BH)-NS systems have taken central stage in astrophysics due to their multi-messenger character and wide-ranging opportunities to advance areas of physics, including gravitational waves, high-density nuclear matter, extreme gravity, and the origin of heavy elements.
The first confirmed BNS merger was a multi-messenger event which manifested itself via a gravitational wave signal~\cite{Abbott17a}, a short gamma-ray burst~\cite{Goldstein17,Savchenko17}, and a range of transients spanning the entire electromagnetic spectrum, from X-rays~\cite{troja17} to radio emissions~\cite{Alexander17}.
The latter, originates from heavy-element nucleosynthesis which
takes place in the ejecta of NS mergers~\cite{Lattimer74}.
The heat from radioactive decay following the rapid neutron capture process (the $r$-process) powers a unique astrophysical transient known as a ``kilonova''.
This transient lasts for about a week, is characterized by rapid reddening, and becomes a thousand times more luminous than a conventional nova; hence its name.
Kilonovae are essentially the $r$-process observed in-the-making with the first one, AT2017gfo, being discovered in coincidence with the first BNS gravitational wave detection, GW170817~\cite{Abbott17b, Abbott17c}. 

Multiple kilonova models have been proposed to interpret the signal of AT2017gfo (see e.g. review by Metzger~\cite{metzger19}).
However, the complexity of the problem precludes definitive conclusions~\cite{korobkin21}.
The most accurate observations of AT2017gfo consist of time-dependent ultraviolet, optical, and infrared spectra~\cite{Chornock2017, Nicholl17}, currently unmatched by theoretical models.
This prevents scientists from answering some of the critical questions about kilonovae, such as: how much $r$-process is produced? 
What is the detailed composition of the ejecta?
Are BNS mergers the main $r$-process sites in the universe?

Most current models assume that the ejecta contains two well-separated components with uniform composition: very neutron-rich ``red'' ejecta, producing the heaviest elements, and moderately neutron-rich ``blue'' ejecta. The colors come from the fact that heavy nuclei include lanthanides and actinides, which have notoriously high opacity in the opital bands~\cite{fontes20}, emitting light mostly in the near infrared. The ``blue'' ejecta is lanthanide-free and may thus produce a short-lived ``blue'' optical transient.
Progress can be made by more comprehensive and advanced models, for example via the inclusion of more complex compositional and morphological elements of the ejecta.
In this project, we go beyond this assumption and simulate more general, non-uniform inhomogeneous expansion of the ejected material in 3D.

There are several mechanisms that can unbind neutron-rich matter from a BNS merger~\cite{rosswog14}.
Here, we focus on the wind ejecta from a post-merger accretion disk~\cite{miller19b}.
Such a disk is left after the central hypermassive merger remnant collapses to a BH.
Visualizing the kilonova helps to gain insights in the ways the morphology and composition are featured in the observed spectra and is a step towards an accurate interpretation and characterization of kilonova spectra for AT2017gfo and future detections. 

Our pipeline incorporates several state-of-the art tools developed at Los Alamos National Laboratory (LANL) to tackle different facets of the problem. 
The most realistic, state-of-the art morphologies for accretion disk wind and ejecta to date have been recently provided by Miller et al.~\cite{miller19b} 
who used the $\nu$bhlight code~\cite{miller19a} to simulate the disk dynamics in a fixed general-relativistic background including Monte-Carlo neutrino transport.
This simulation required approximately 400 cores for 3 weeks of walltime. As a hybrid simulation, it used over one million finite volume zones, two million tracer particles,
and ten million Monte Carlo packets.
An example of the results can be seen in  
Figure~\ref{fig:ye-ballistic}.

Kilonova ejecta expansion over many scales is then simulated with FleCSPH~\cite{loiseau20}, which is a smoothed particle hydrodynamics (SPH) code.
SPH is an explicit numerical mesh-free Lagrangian method designed to solve the equations of hydrodynamics by representing the fluid with a set of moving and interacting particles~\cite{rosswog15}.
Each particle is equipped with a smoothing length and a kernel, measuring the range and intensity of hydrodynamic interactions with other particles, respectively.
SPH possesses excellent conservational properties, adaptive resolution across multiple scales, and natural treatment of vacuum.
The FleCSPH code is based on FleCSI~\cite{charest17}---a compile-time configurable framework, which supports parallel and distributed computing for multiple data topologies, such as mesh, n-array, or ntree.
FleCSPH's ntree is a parallel binary, quad, and octree used for neighbor search and gravitational interactions in one, two, and three spatial dimensions, respectively. 
Features currently implemented in FleCSPH include fast multipole method for long-range gravitational interactions~\cite{korobkin21b}, variety of effective potentials, tabulated equation of state (EOS), tabulated radioactive heating source, material strength, and both Newtonian and relativistic SPH~\cite{tsao21}, making it perfectly suitable for the task.

\section{Astrophysical model}
\label{sec:astophyiscal_model}

In this work, we focus on the expanding ejecta and the kilonova aspect of a NS merger.
It is believed that in most cases the heavy remnant of the merger collapses to a BH, leaving behind an accretion disk~\cite{metzger19}.
The latter ejects neutron-rich material which becomes a production site for $r$-process elements. 
It has recently been demonstrated that the morphology of the ejecta plays a substantial role in shaping the spectra and light curves of the kilonova~\cite{korobkin21, hainzel21}.
We captured the output flux from the accretion disk simulations with $\nu$bhlight~\cite{miller19a} as initial data for the kilonova expansion.

\begin{figure}[htb]
  \includegraphics[width=\linewidth]{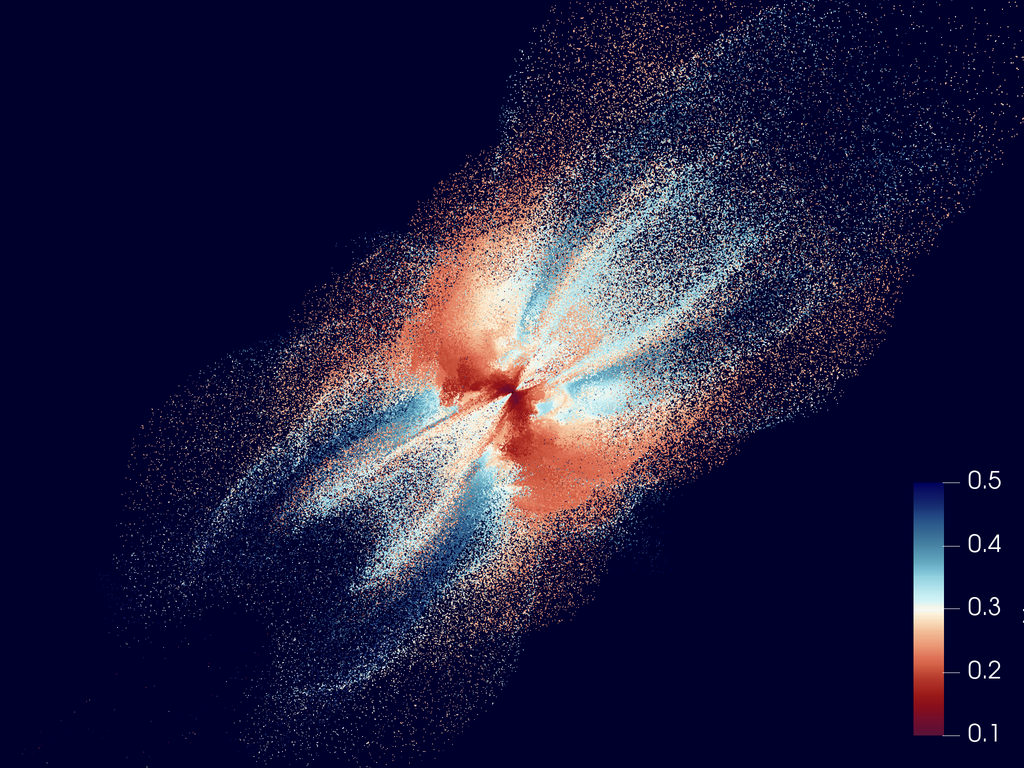}
  \caption{
    Initial distribution of the electron fraction (number of electrons per nucleon) for 
    accretion disk ejecta in velocity space~\cite{miller19b}.
    The velocity is sampled at the times when particles exit the outflow boundary of 
    the spherical grid. The ejecta is discretized with 27M particles.
    The red color corresponds to high neutron 
    richness with strong $r$-process nucleosynthesis.
    The blue color indicates high electron fraction and lower opacity, marking 
    regions that are potentially responsible for the bright optical emission and the ``blue'' kilonova.
  }
  \label{fig:ye-ballistic}
\end{figure}

\begin{figure}[htb]
  \includegraphics[width=\linewidth]{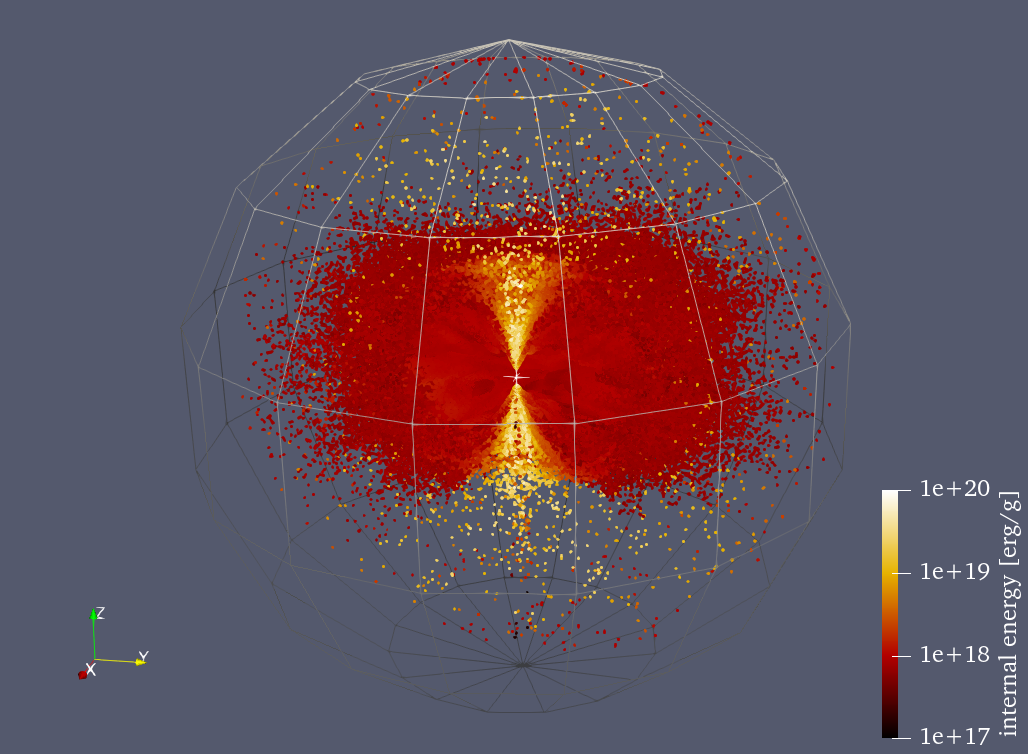}
  \caption{
    Sample initial compactified particle configuration, color-coded 
    according to the internal energy that particles will have when they cross the 
    boundary.
    Inside the spherical region, the particles move radially according to the 
    formula (\ref{eq:inversion}).
    When the particles cross the boundary and enter the SPH domain, their density is in 
    accordance with the simulated outgoing flux from the grid-based code {\tt $\nu$bhlight},
    and other physical quantities are interpolated from the grid.
    Thus, the outer boundary of the grid-based code serves as the inner flux boundary of 
    the SPH code.
  }
  \label{fig:uint_initial}
\end{figure}

The expansion of the kilonova covers multiple orders of magnitude in scale, density, and temperature.
With that, a Lagrangian particle method, such as SPH, is better suited than a grid-based code to model this phase. 
To map the ejecta data from $\nu$bhlight into FleCSPH, we 
sample the outgoing flux from the accretion disk wind simulation and convert it into particles using
the Wendland $C_6$ kernel for the SPH interpolation (see reviews of the SPH method for details~\cite{rosswog15}).
We discretize the flux emerging from the spherical boundary using regular time intervals $\Delta t$.
The total mass of the ejecta $m_{\rm ej}$ is computed by integrating the flux over the entire accretion disk simulation time (about 1 second).
Assuming the ejecta will be represented by $N_p$ particles of equal mass $m_p = m_{\rm ej}/N_p$, we estimate how many particles will emerge from the boundary during $\Delta t$, and then randomly sample that many particles using the normalized flux as the probability distribution over the sphere.
To avoid unphysical inwardly directed pressure gradients at the inflow boundary, we prepare the compactified configuration of all the particles inside the spherical inflow boundary using an inversion map.
During the simulation, particles inside the sphere are advected using the time-dependent inversion, passing through the boundary at times such that their emergent configuration correctly mimics the outgoing flux.
The equation of advection for a particle $a$ inside the initial sphere (with radius $R_{\rm in}$) is:
\begin{equation}
{r_a(t) = \frac{R_{\rm in}}{1 + v_r (t - t_{\rm ex})/R_{\rm in}}}
\label{eq:inversion}
\end{equation}
where $r_a$ is the radial distance of the particle from the origin, $v_r$ is its radial velocity upon exit, and $t_{\rm ex}$ is the expected exit time.
Because the density in our SPH formulation is computed using the local density of particles, this setup provides continuity of the flux and ensures accurate pressure gradients at the boundary.
As soon as particles exit the injection sphere ({$R_{\rm in} = 3500$~km} in our case), they undergo regular hydrodynamic evolution.
Figure~\ref{fig:uint_initial} demonstrates this initial compactified particle distribution.

Since $r$-process nucleosynthesis happens on timescales of a second and it is generally very challenging to include an inline $r$-process network in hydrodynamic simulations, we apply precomputed radioactive heating contributions.
The latter are given in form of tables and were computed with WinNet~\cite{korobkin12} using the Helmholtz EOS~\cite{timmes99} and asumming that $40\%$ of the generated energy escapes in the form of neutrinos. 
For validity checks, we also compared with results from SkyNet~\cite{lippuner17} and PRISM~\cite{mumpower18} for a few representative trajectories. 

At early times of kilonova expansion, the internal energy of the outflow is heavily dominated by radiation, followed by a subdominant contribution from electrons and an almost negligible contribution from ions.
For this reason, we find it sufficient to use the ideal fluid equation of state $P = (\Gamma-1)\rho\varepsilon$, with $P$, $\rho$, and $\varepsilon$ representing the pressure, density, and specific internal energy, respectively, and $\Gamma=4/3$.
At early times, the optical depth of the outflow is extremely high, such that radiation is fully trapped.
Simple estimates for density and temperature of the ejecta show that throughout the morphology-forming times (the first minute) the contribution to energy and pressure from ions is completely negligible. 
The electron contribution can be neglected as well as it is heavily subdominant.
Contributions from electron-positron pairs can be estimated following e.g.~\cite{farouqi10}, but would only make significant differences at high temperatures which exceed $5\times 10^9$~K. 
This is not the case in our simulations. 

The ejecta is expanded in a static gravitational potential of a central point mass, mimicking the $2.58$-solar-mass BH from the original accretion disk simulation, with added softening of the $1/r$ potential at a length scale of $\varepsilon = 10^3\;{\rm km}$.
This simulation used 128 cores for 48 hours.
After about 30 seconds of simulation time, some of the particles form dynamic fallback.
Since, in this work, we are interested in the expanding part of the flow with progressively larger timesteps and sizes, we truncate these particles to avoid complications due to potentially small timesteps near the origin.
A similar method was employed in Rosswog et al.~\cite{rosswog14}.
Nuclear heating and gravity reshape the ejecta: the former makes it more isotropic~\cite{rosswog15}, while the latter slows it down and takes up some of the material, forming dynamical fallback. 
Without these two factors, the output flux would just expand as shown in Figure~\ref{fig:ye-ballistic}.
The figure shows the initial electron fraction in the ejecta, with red indicating more neutron-rich and blue less neutron-rich material.
The ``red'' ejecta would subsequently synthesize lanthanides while the ``blue'' part would be lanthanide-free and produce the ``blue'' kilonova.
As can be seen, there is a broad equatorial belt of very neutron-rich material, while the polar and medium-latitude regions are occupied by the ``blue'' component.

\section{Visualization}
\label{sec:visualization}

ParaView offers several options to visualize particles from a SPH simulation. 
One example is the \texttt{Surface} representation which we can use to obtain a quick overview of the simulation results. 
We can also apply either the \texttt{Point Gaussian} representation or the \texttt{Glyph} filter to plot particles as spheres with sizes defined by the smoothing length. 
However, since the objective of SPH is to represent the continuum with particles, we are most interested in applying volume rendering to our simulation results. 

\subsection{Volume rendering of SPH particles}
\begin{figure}
    \centering
    \begin{tabular}{cc}
         \includegraphics[width=0.49\linewidth]{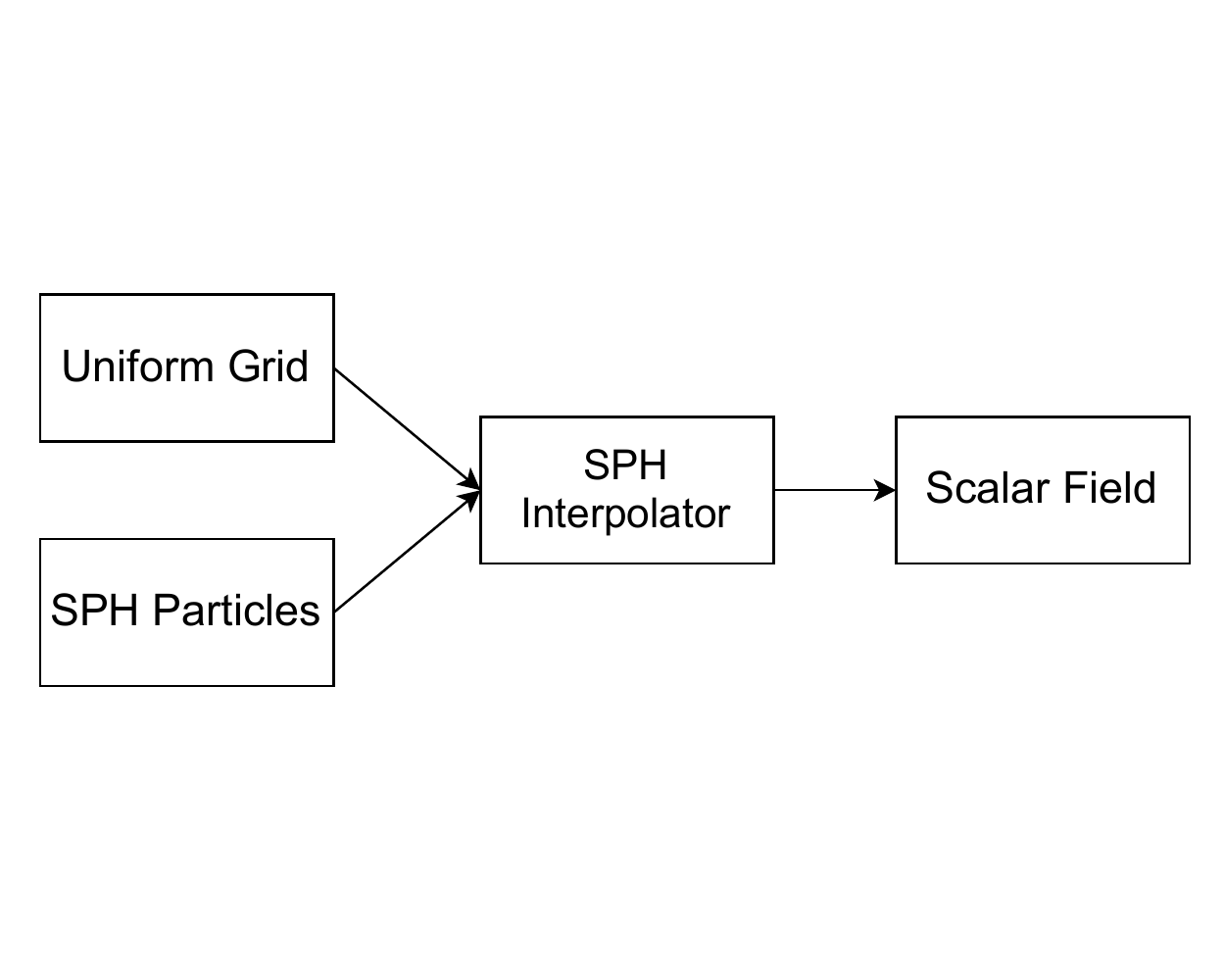} &  
         \includegraphics[width=0.49\linewidth]{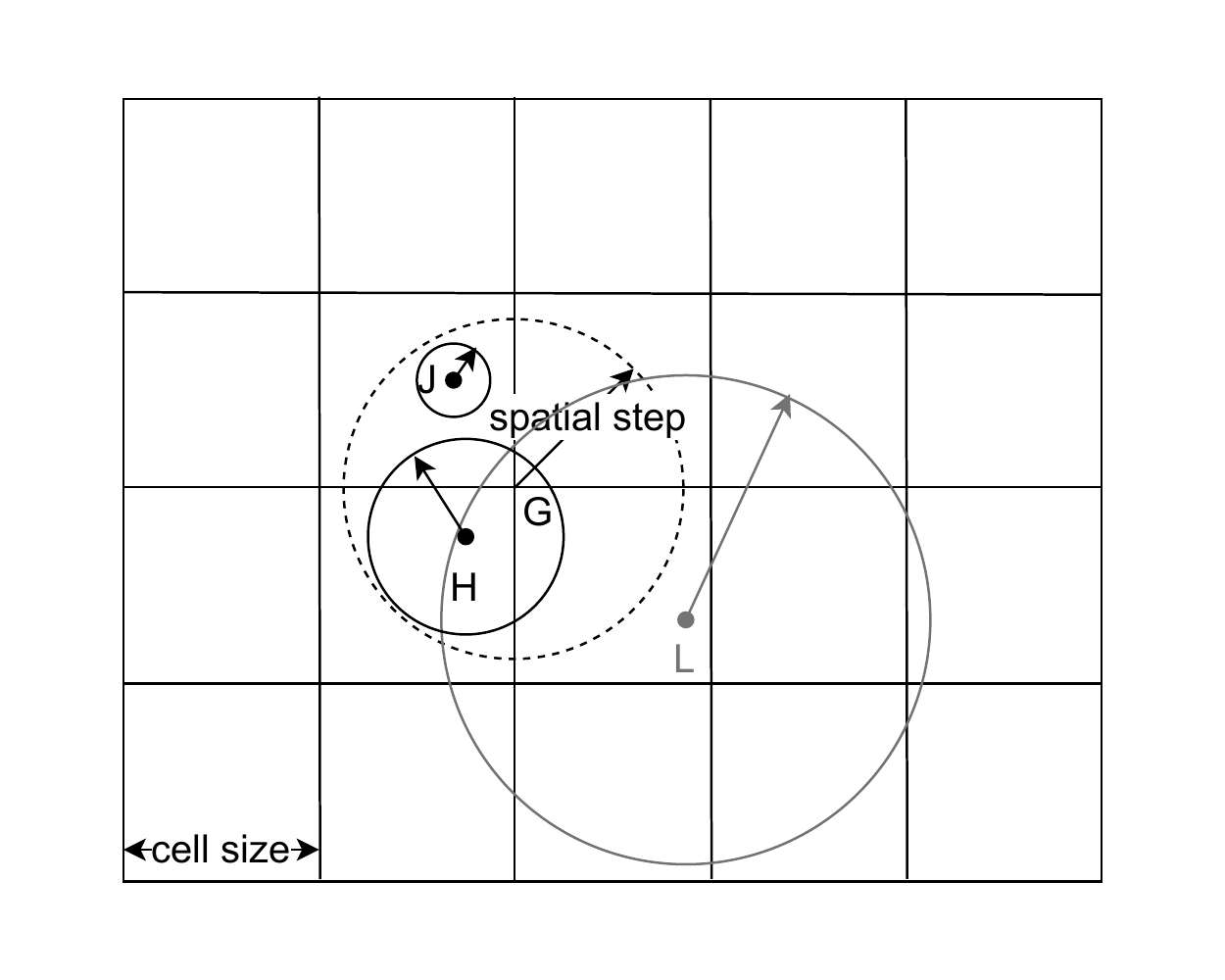} \\
         a & b
    \end{tabular}
    \caption{Left: ParaView SPH Volume Interpolator pipeline. Right: An example for uniform grid interpolation of data from SPH particles with varying smoothing radius, i.e. size, represented by circles.}
    \label{fig:sph_interpolation}
\end{figure}

Paraview includes a \texttt{SPH Volume Interpolation} filter to turn particles into a uniform grid which can then be visualized with volume rendering. 
As shown in Figure~\ref{fig:sph_interpolation}a, the Interpolator takes two input datasets, a uniform grid provided by the \texttt{vtkBoundedVolumeSource} and the SPH particles. Through the graphical user interface (GUI), a user can set the bounding box of the particle dataset and configure the ${xyz}$-dimensions, origin and spacing for the uniform grid. A user can also specify various aspects of the SPH particles, for example, arrays for particle mass, density, and cutoff radius (proportional to smoothing length) as well as the SPH smoothing kernel to match the one used in the simulation. At first sight, this filter seemed to be the perfect tool for our work.  
However, we soon encountered several major difficulties which made it unsuitable for our purposes.

The first issue lies in the design of the current ParaView pipeline architecture. 
Although ParaView provides a widget for users to interactively specify the bounding box of the particle dataset, the Interpolator cannot automatically calculate the bounding box from its particles input and use the result to configure the \texttt{vtkBoundedVolumeSource} input.
This would require \emph{information} to flow in the inverse direction of the pipleline. 
Over the course of our simulation, the accretion disk ejecta expands from the size of a small planet to the size of the solar system. The lack of automatic adjustment of the bounding box poses a major difficulty. 
We were therefore required to use an external interpolation to manually compute and set the bounding box through the GUI when producing animations. 

The second and more severe issue lies in the implementation of the interpolation algorithm in the \texttt{SPH Volume Interpolation} filter. Here, the Interpolator uses a user-specified parameter called \texttt{spatial step} to accelerate the search of particles. 
As shown in Figure~\ref{fig:sph_interpolation}b, for each grid point \textit{G} in the uniform grid, the Interpolator performs a spatial range search within the radius of \texttt{spatial step}. It then iterates through all the particles \emph{within} the radius, calculates, and accumulates contributions from each of them.
This approach works well when particle densities and thus cutoff radii don't vary by a large amount. 
However, in our case they span several orders of magnitude.
Using a single fixed search radius therefore leads to inaccurate interpolation results.
In our simulation, particles near the center of the ejecta have much higher densities compared to the particles at the periphery. 
In lower-density regions, particles are also spaced more sparsely. 
As depicted in Figure~\ref{fig:sph_interpolation}b, a uniform \texttt{spatial step} will be able to find the smaller, higher density particles \textit{H} and \textit{J} near the grid point \textit{G} but will miss the larger but lower density particle \textit{L} whose center is outside of the search radius and completely ignores the contribution from \textit{L} to \textit{G}. As a result, particles in the outer regions of the ejecta are mostly unaccounted for and are \textit{invisible} in the volume rendered image.

We also encountered some further unidentified issues with the \texttt{SPH Volume Interpolator}, resulting in some un-physical interpolation. 
For example, the electron fraction $Y_e$ is physically constrained in the range of $[0, 0.5]$. 
However, the Interplator sometimes gives us either negative values or values much larger than 1. 
All the described difficulties prompted us to implement our own SPH particle-to-mesh interpolator which we will describe in the following. 
Fortunately, it is straightforward to generalize the standard SPH interpolation formula so that it can be applied to grid points~\cite{rosswog15}.
For a physical quantity $F_b$ specified on particles, the interpolated value $F_{ijk}$ can be computed using the following expression:
\begin{align}
    F_{ijk} &=\sum_{b\in O(\vec{r}_{ijk})} F_b \frac{m_b}{\rho_b} W(|\vec{r}_{ijk} -\vec{r}_b|, h_{ijk}),
    \label{eq:sph-interp}
\end{align}
where $\rho_b$ and $m_b$ are the density and the mass of the particle $b$, $h_{ijk}$ is the ``smoothing length'' for the grid point $\vec{r}_{ijk}$, $O(\vec{r}_{ijk})$ is the $h_{ijk}$-neighborhood of $\vec{r}_{ijk}$, and $W(r,h)$ is the SPH kernel.
To apply these formulae, we need to first compute the smoothing length $h_{ijk}$ for each grid point. 
Our interpolation algorithm is therefore performed in two passes over the particles. In the first pass, we determine the smoothing length on a grid by simply borrowing the smoothing length of the nearest particle for each grid point.
For each particle, we traverse the grid points within its interaction radius and update their nearest particle information.
Taking Figure~\ref{fig:sph_interpolation}b as example, the algorithm first marks \textit{L} as the nearest particle of grid point \textit{G} and records the distance ${|G - L|}$. This is later updated to particle \textit{H} and the distance ${|G - H|}$. 
The first pass ensures that valid contributions from distant particles are accounted for.
In the second pass, we use expression~(\ref{eq:sph-interp}) with the known smoothing length on the grid to compute values of the physical quantities.

Even with properly interpolated grid data, we still faced severe numerical precision issues with the volume renderer in ParaView. Our simulations adopt the centimetre–gram–second system of units (CGS) which are widely used in the astrophysics community. 
Since we are simulating astronomical events, many of our data values are also astronomical. For example, the $xyz$-coordinates might range from $-10^{10}$ to $10^{10}$~cm. At any reasonable mesh resolution, we will have a cell size (called \emph{spacing} in ParaView) in the order of $10^8$. ParaView seems to have difficulty propagating rays along such \emph{astronomically large} cells and shows either nothing or all black pixels as a result. As suggested by Kitware Inc., we scaled the spacing for the grid to be in the units of astronomical unit (AU) which seems to resolve the numerical precision issue.

\subsection{Physics-based transfer function}
\begin{figure}
    \centering
    \includegraphics[width=\linewidth]{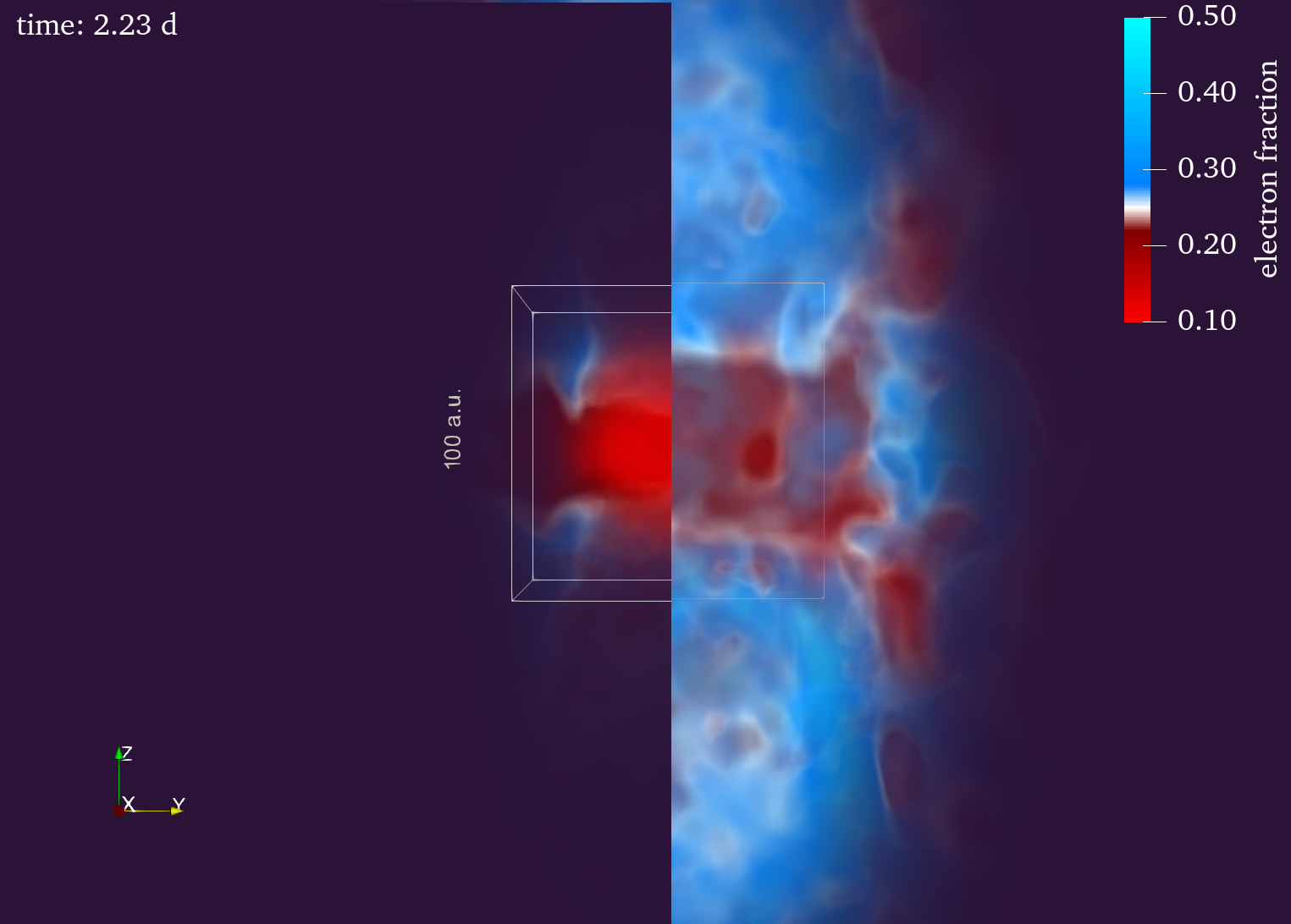}
    \caption{Volume rendering of the electron fraction using two different opacity transfer functions: simple density (left) vs the physics-based absorptance (right) from Eq.(\ref{eq:Abs}). On the left, the opacity is strongly underestimated, creating an image of the ejecta that is almost completely transparent.
    The right half shows correct physical picture for this epoch (2.23~days) with complex structures in the electron fraction.
    }
    \label{fig:physical_opacity}
\end{figure}

The main goal of this project is to answer the question: \emph{what would a kilonova realistically look like up close?} 
To answer this, we need an opacity transfer function that is physics-based to apply in our volume rendering. 
In general, both the color and opacity transfer functions depend on the electron fraction $Y_e$.
Regions with low $Y_e$ correspond to a neutron-rich environment where the heaviest elements, including lanthanides and actinides, are synthesized.
Due to half-filled atomic $f$-orbitals, the latter possess an extremely complex level structure with tens of thousands of energy levels and up to a few million lines in the optical bands.
As a consequence, ejecta that contains lanthanides and actinides is associated with a dense line blanketing of the bound-bound opacity in the optical range of the spectrum~\cite{fontes20}.
Lanthanide-free material, on the other hand, is less opaque in the optical and appears as ``blue'' to the observers.
With that, it is convenient to adopt a red-blue colormap on $Y_e$ and use red color to indicate more neutron-rich material where lanthanides are present, and blue color to show less neutron-rich and lanthanide-free matter. 
We map the physical opacity to the absorptance of the grid cells using the following formula:
\begin{align}
    A = 1 - e^{-\tau},
    \label{eq:Abs}
\end{align}
where $\tau$ is the optical depth across the cell, estimated as $\tau = \varkappa\ \rho\ \Delta x.$
Here, $\rho$ is the density in the grid cell, $\Delta x$ is its size, and $\varkappa$ is the material opacity. 
In this work, we use $\varkappa=10\ \cmg$ for lanthanide-rich ejecta, and $\varkappa=1\ \cmg$ for the lighter $r$-process, lanthanide-free ejecta, which is a simple approximation that has been successfully applied for constructing kilonova models in the past~\cite{grossman14}.
An electron fraction of $Y_e=0.25$ is used as threshold value between lanthanide-rich and lanthanide-free ejecta, similar to previous studies~\cite{korobkin12}.
Figure~\ref{fig:physical_opacity} shows the effectiveness of such an opacity transfer function to reveal the complex structure of lanthanide-rich material responsible for the blue kilonova, in comparison with the na\"ive use of density in that role. 
Because density spans several orders of magnitude, when its normalized value is used as the opacity transfer function, it looks significantly more transparent and redder than expected. 
\section{Results}

\begin{figure*}[htb]
  \centering
  \begin{tabular}{cc}
  \includegraphics[width=0.48\linewidth]{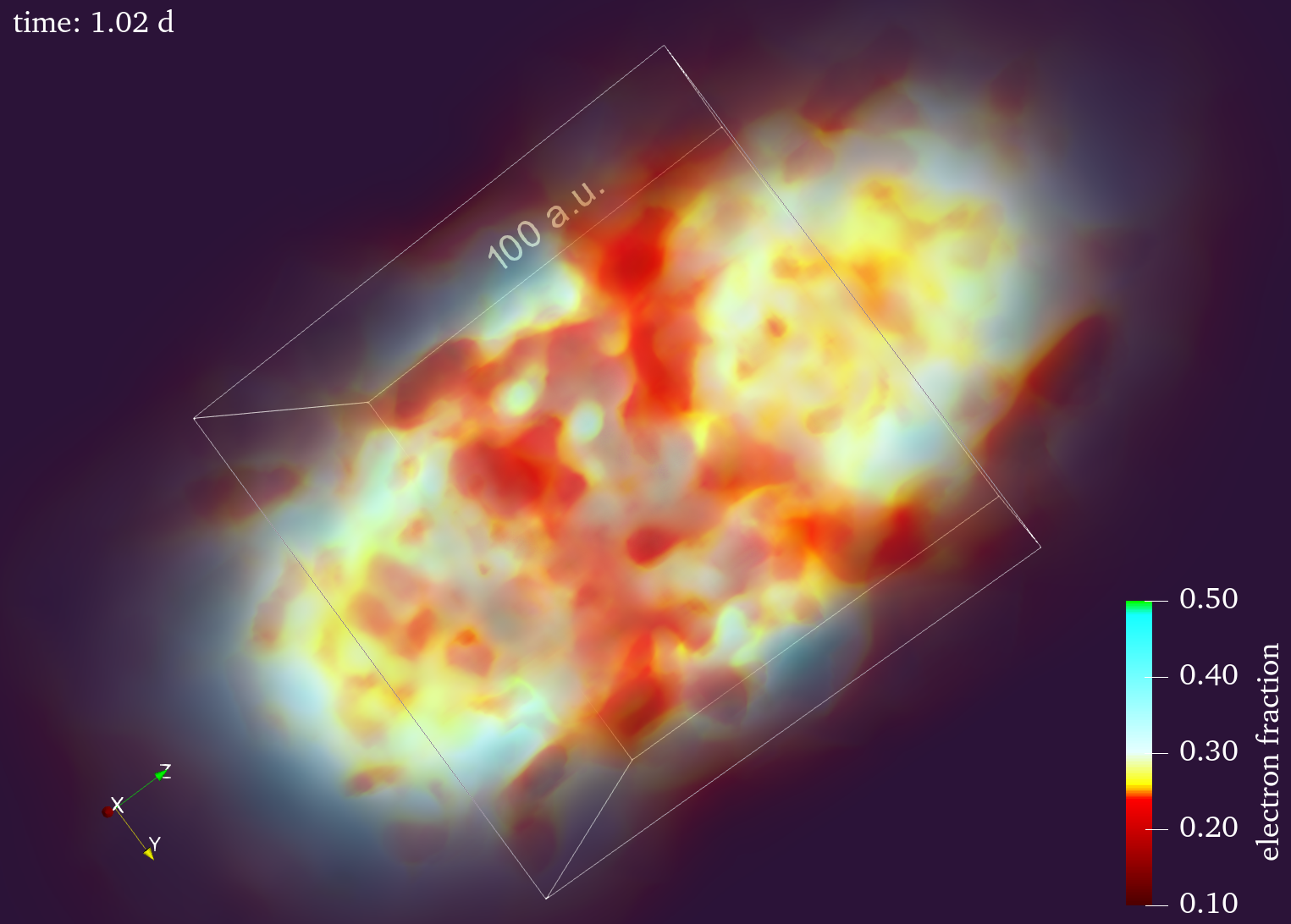} &
  \includegraphics[width=0.48\linewidth]{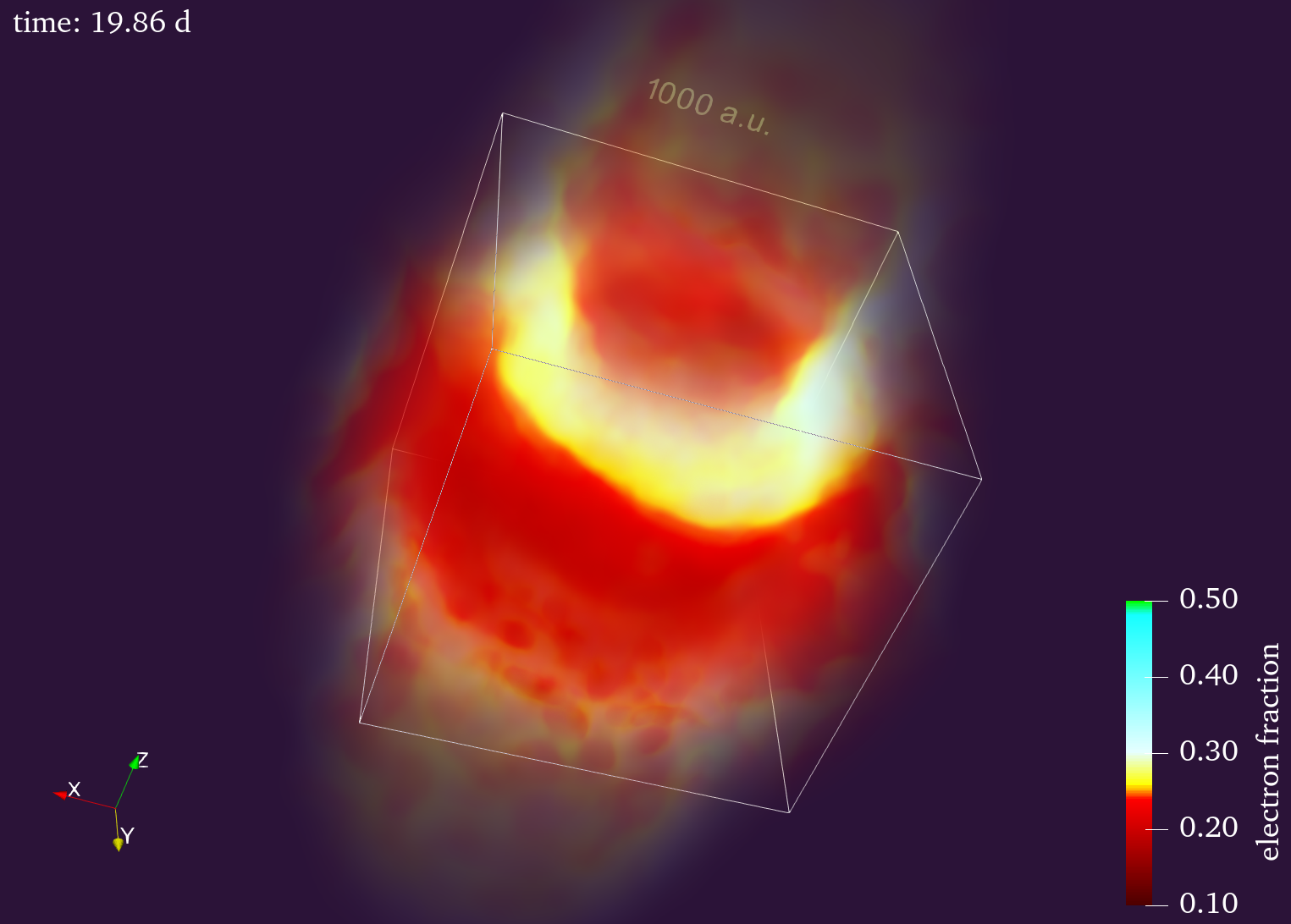}
  \end{tabular}
  \caption{
    Left: kilonova rendering at $t=1$~day.
    Right: same at $t=20$~days.
    Opacity transfer function is computed according to Eq.(\ref{eq:Abs}), while the color 
    maps out distribution of the electron fraction.
  }
  \label{fig:kn-vis}
\end{figure*}
Our most realistic visualization of the kilonova is given in Figure~\ref{fig:kn-vis}.
The left panel shows the ejecta at the epoch $t=1$~day, while the right panel illustrates the view at $t=20$~days.

At day 1, we can see various "blob-like" structures of lanthanide-rich (red) material.
They are mostly located around the central equatorial region of the kilonova but also reach into the polar lobes. 
These blobs are seeded by turbulence in the accretion disk from the magneto-rotational instabilities.
The inhomogeneities are then exaggerated in the disk wind.
Even in artificial colors and the presence of red material blobs, the kilonova looks blue at this point in time. 
By day 20 the ejecta expanded by one order of magnitude in size and we can see a significant color change from blue to red. 
As the kilonova expands, a photospheric recession advances the photosphere into the core uncovering more lanthanide-rich ``red'' material.
Our visualization thus finds a rich structure of regions where the lighter $r$-process takes place.
For once, there no is simple concentration e.g. only in the polar lobes; instead regions span a band in the middle latitudes, which would show up in observations as a ``blue'' kilonova for some orientations.

\section{Conclusion}
\label{sec:conclusion}

We have produced, for the first time, a physically realistic rendering of a kilonova, resulting from the coalescence of a neutron star binary.
The realism of our model is based on the fact that we used a state-of-the art model for the accretion disk wind, realistic nucleosynthesis, plausible models for physical opacities, and correctly expanded hydrodynamical flow.
We use flux compactification to interface the grid-based hydrodynamics with the particle-based one (Section~\ref{sec:astophyiscal_model}).
For visualization, the expanded flow is interpolated on the grid, and the inhomogeneous physical opacity is mapped to the absorptance using Eq.(\ref{eq:Abs}).
The final result, a synthesized image of the disk wind ejecta, illustrates the complexity of the kilonova structure. 
The nature of the rich features and their influence on the observed kilonova spectra will need further investigation.
Current astrophysical models use simplistic assumptions such as spherical symmetry and single- or two-component morphology.
Our work clearly demonstrates how little may actually be captured by such simple models.
Future work may include more advanced mapping between the electron fraction and the opacity ($\kappa(Y_e)$). 
Furthermore, the ejecta and associated kilonova are expected to differ depending on the final outcome of the NSM. 
Depending on the neutron stars' masses, the result of the merger can be a stable neutron star, a short-lived hypermassive neutron star, or a black hole. 
Visualization and analysis of such ejecta data and kilonova for different merger configurations would be an interesting future undertaking. 

During our work, we encountered several issues when visualizing SPH particle data with highly variable smoothing length in ParaView.
Tracking down the origin of the problem, we give a suggestion for mapping SPH data onto a grid in order to take full advantage of Paraview's volume rendering capabilities.

\section*{Acknowledgments}
The authors thank Ryan T. Wollaeger and Soumi De for their helpful feedback.
Funding for this project was provided by LANL Ristra Project and Laboratory Directed Research and Development (LDRD). 
Alexandra Stewart would like to thank LANL's Data Science at Scale Summer School for the opprotunity to work on this project. 
This work used resources provided by the LANL Institutional Computing Program. 
LANL is operated by Triad National Security, LLC, for the National Nuclear Security Administration of the U.S. DOE (Contract No. 89233218CNA000001).
This work is authorized for unlimited release under LA-UR-21-28322.


\end{document}